\documentclass[10pt, conference, letterpaper]{IEEEtran}
\usepackage{mystyle}

\newcommand{\name}{\texttt{Chocolatine}}

\newcommand{\etal}{\textit{et al.}~}

\providecommand{\ie}{\emph{i.e.,} }
\providecommand{\eg}{\emph{e.g.,} }

\begin{document}

\title{Chocolatine: Outage Detection for Internet Background Radiation}

%

\author{
	Andreas Guillot{$^{\ast}$}, Romain Fontugne{$^{\dag}$}, Philipp
    Winter{$^{\ddag}$}, Pascal Merindol{$^{\ast}$}\\Alistair King{$^{\ddag}$}, Alberto Dainotti{$^{\ddag}$}, Cristel Pelsser{$^{\ast}$}\\
	$\ast$ICube, University of Strasbourg -- France ~~~~$\dag$IIJ Research Lab
    -- Japan ~~~~$\ddag$CAIDA, UC San
    Diego -- USA\\
}

\maketitle              

\begin{abstract}
The Internet is a complex ecosystem composed of thousands of Autonomous Systems
(ASs) operated by independent organizations; each AS having a very limited view
outside its own network. These complexities and limitations impede network
operators to finely pinpoint the causes of service degradation or disruption
when the problem lies outside of their network. In this paper, we present \name,
a solution to detect remote connectivity loss using Internet Background
Radiation (IBR) through a simple and efficient method.  IBR is unidirectional
unsolicited Internet traffic, which is easily observed by monitoring unused
address space.  IBR features two remarkable properties: it is originated
worldwide, across diverse ASs, and it is incessant. We show that the number of
IP addresses observed from an AS or a geographical area follows a periodic
pattern. Then, using Seasonal ARIMA to statistically model IBR data, we predict
the number of IPs for the next time window. Significant
deviations from these predictions indicate an outage. We evaluated \name~using
data from the UCSD Network Telescope, operated by CAIDA, with a set of
documented outages. Our experiments show that the proposed methodology achieves
a good trade-off between true-positive rate ($90$\%) and false-positive
rate ($2$\%) and largely outperforms CAIDA's own IBR-based detection method.
Furthermore, performing a comparison against other methods, i.e., with BGP monitoring and
active probing, we observe that \name~shares a large common set of outages with them in addition to many specific
outages that would otherwise go undetected.
\end{abstract}

\begin{IEEEkeywords}
Outage detection, Internet Background Radiation, ARIMA
\end{IEEEkeywords}

\section{Introduction}

Connectivity disruptions caused by physical outages, software bugs,
misconfiguration, censorship, or malicious activity, occur repeatedly on the
Internet~\cite{aceto2018comprehensive}. Monitoring the state of Internet
connectivity is useful to raise public awareness on events of
intentional disconnection due to censorship~\cite{hall-censorship:2018}. It further helps
operators pinpoint the location of an outage, \ie the place where there is a
loss of connectivity,
when it happens outside their reach. This enables to speed up recovery as the
correct network operator team can be contacted directly instead of reaching out
to the global network operators community via mailing lists or personal
contacts. Fast outage detection is also useful to locally switch
to backup routes, when available~\cite{Holterbach:NSDI2019}.

A few methods exist to detect connectivity outages. Monitoring for withdrawals
of BGP prefixes is a commonly used approach, but it can only observe outages
that affect the control plane~\cite{dyn,akamai}.  Data-plane approaches solve
this problem, and can be either based on active measurements---\eg
Trinocular~\cite{quan2013trinocular} sends pings to 4\,M remote /24 address
blocks to measure their liveness---or on passive traffic
analysis---Disco~\cite{shah2017disco} relies on the long-running TCP connections
between RIPE Atlas probes and their controlling infrastructure to identify
bursts of disconnections.

Another data-plane approach for the detection of connectivity outages,
is based
on the analysis of Internet Background Radiation (IBR)~\cite{DainottiCCR2012}.
IBR is unsolicited traffic captured by darknets (also known as network
telescopes), which announce unused IP prefixes on BGP, \ie there are no actual services
running in the prefix, nor ``eyeballs''. IBR is composed of a constantly evolving mix of
various phenomena: network scans, the results of malware infections, DoS attacks
using spoofed IPs from the range announced by the
telescope~\cite{benson2015leveraging}, packets from misconfigured (or with a
polluted DHT) BitTorrent clients, etc.~\cite{wustrow2010internet}.  By
leveraging the pervasiveness of IBR sources, and the consistent presence of
traffic, we can infer a connectivity outage for a given geographic area or
Autonomous System (AS) based on a significant reduction of IBR traffic
that originates from them.  In addition, Dainotti et
al.~\cite{dainotti2011analysis,DainottiCCR2012} demonstrated that IBR can effectively complement both
control-plane and active probing data-plane approaches: both in terms of
coverage (not all networks respond to pings) and in terms of information that it
provides (\eg confirming outbound connectivity for a remote network even when
inbound connectivity is disrupted).

The IODA system from CAIDA~\cite{ioda} has recently operationalized this method
for extracting time series, \ie ``signals'', at different spatial grain (\eg
countries or ASs). However, IODA's current automated detection algorithm is
simplistic (a threshold based on the last 7 days moving median) and unable to
take into account the IBR's noise and the intensity variability of the signal.
Indeed, in order to avoid an overwhelming amount of false positives, the
threshold is currently set to raise an outage alert when the signal intensity
drops under 25\% of the intensity of the median value observed in the last 7
days. That is, an outage is detected only when there is a severe connectivity
loss, leaving many cases of connectivity loss undetected~\cite{ioda-help}. In
particular, the test remains the same whatever the period of the day and the
week, such that a drop occurring in an usually busy period is treated the same
as if it was occurring during an inactive one. In one word, this naive model is
static, and as such, challenging to calibrate, as it does not take into account
any trends in the traffic.

In this work, we take these trends into account by applying Seasonal ARIMA
(SARIMA)~\cite{gooijer:ijf06}, a popular technique that forecasts the behavior
of the time series extracted at the UCSD Network Telescope~\cite{caida-telescope}.
More specifically, we analyze the number of unique source IP addresses that
try to reach the darknet of different countries/ASs.
\name~is sensitive and robust, respectively to the seasonality and noise observed in the
data. We show that it is able to detect outages with a true positive rate of
$90\%$ and a false positive rate of $2\%$ with a detection delay of only 5
minutes.
Additionally, the comparison with CAIDA's method showed that \name~can detect a large share of outages seen by other data sources, as well as additional specific outages.
Another benefit of \name~is that its algorithm automatically
self-tunes on time series exhibiting very different magnitudes and levels of
noise (\eg time series of IBR extracted for ASs and countries of different sizes
and with different compositions of IBR-generating sources).
As a result, \name~can be applicable to other seasonal and
noisy data sources related to Internet traffic activity.

The remainder of the paper is structured as follows: background on main outage
detection methods is first provided in Section~\ref{sec:back}. In
Section~\ref{sec:data}, we then introduce the dataset we use, and explain why it
is suited for outage detection. In Section~\ref{sec:method}, we describe \name's
high-level design. We also illustrate our outage detection process with a case
study of censorship occurred during the Egyptian revolution
(Section~\ref{sec:case-studies}). In Section~\ref{sec:validation}, we evaluate
\name, validating it with ground truth data and also comparing its performances
against several current outage detection algorithms.  Lastly, we address the
reproducibility of our experiments in Section~\ref{sec:repro}.

\section{Background}\label{sec:back}

Outage detection can be achieved with different measurement techniques
and performance indicators. A recent survey~\cite{aceto2018comprehensive} 
provides a taxonomy of most existing techniques, including three main monitoring categories:
\textit{active}, \textit{passive}, and \textit{hybrid}.
We reuse this terminology here.

Active monitoring techniques generate traffic in order to collect
information and examine the state of networks.
Most active monitoring approaches are based on variants of \textit{ping} and
\textit{traceroute}, and rely on a set of \textit{vantage points}
(\ie the devices that perform the measurements) that are usually
distributed across different networks.
For example, RIPE Atlas~\cite{ripeatlas} is a popular platform for network measurement that is composed
of over 10,000 probes. In \cite{FontugneIMC2018}, Fontugne \etal detect
significant link delay changes and rerouting from RIPE Atlas built-in
measurements. Dasu~\cite{sanchez2014dasu}, on the other hand, is more versatile
than RIPE Atlas. It has been used for diverse measurements, such as broadband
performance measurements, as well as the mapping of the Google CDN infrastructure.
Thunderping~\cite{schulman2011pingin} measures the
connectivity of residential Internet hosts before, during, and after forecast
periods of severe weather.

Passive monitoring techniques collect existing traffic and infer the state of
networks from it. Generally speaking,
they analyze real-user traffic to be close to the user experience. It ensures that the
inferred statistics correspond to real traffic, thus granting a view of a
network's current state.
Different datasets have been leveraged for passive analysis, such as CDN
traces~\cite{RichterIMC2018}, or darknets~\cite{benson2013gaining}.

Outage detection methods also rely on different theoretical modeling techniques
to discriminate outages from normal network conditions.
Trinocular \cite{quan2013trinocular} leverages Bayesian inference to estimate
the reachability of /24 subnetworks.
Disco \cite{shah2017disco} detects surge of Atlas probe disconnections using a
burst modeling algorithm.
Using also Atlas data, authors of \cite{FontugneIMC2018} rely on the central
limit theorem to model usual Internet delays and identify network disruptions.

In this work, we rely on passive measurements collected from CAIDA's network
telescope~\cite{caida-telescope} and employ SARIMA models to forecast IBR time
series and detect outages.

\section{Dataset}\label{sec:data}

The data used for this study is obtained from the UCSD network telescope~\cite{caida-telescope}.
The goal of this section is to provide an overview of the characteristics of
this dataset, and to motivate why it is suitable for outage detection.

The collected data consists exclusively of unsolicited traffic caused by both benign
and malicious activities.
For instance, software and hardware errors, such as bit-flipping or hard-coded IP
addresses, result in IBR traffic.
Network scans and backscatter traffic are another common source of IBR traffic.
Backscatter traffic is usually the consequence of malicious spoofed traffic
sent to a victim and whose replies are returned to unused addresses monitored by
the network telescope.
Consequently, IBR data has been extensively used to study
worms~\cite{wang2011darknet}, virus propagation~\cite{harder2006observing},
and Distributed Denial of Service (DDoS) attacks~\cite{fachkha2014fingerprinting}.

CAIDA's IODA~\cite{ioda} aggregates UCSD network telescope data geographically
and topologically, respectively using
NetAcuity~\cite{netacuity} IP geolocation datasets and longest prefix
matching against BGP announcements from public BGP data~\cite{caida-prefix2as}.
Consequently, we obtain IBR streams per country, regional area (\eg states in
the US, provinces in France, etc.), and AS.
IODA also pre-filters the traffic that reaches the telescope,
removing large components of potentially spoofed-source-IP traffic (since their
presence would significantly alter inference about originating ASs and
geographical areas) using a set of heuristics derived
semi-manually~\cite{dainotti2013estimating}.

\begin{figure*}[h]
    \centering
    \includegraphics[width=\textwidth]{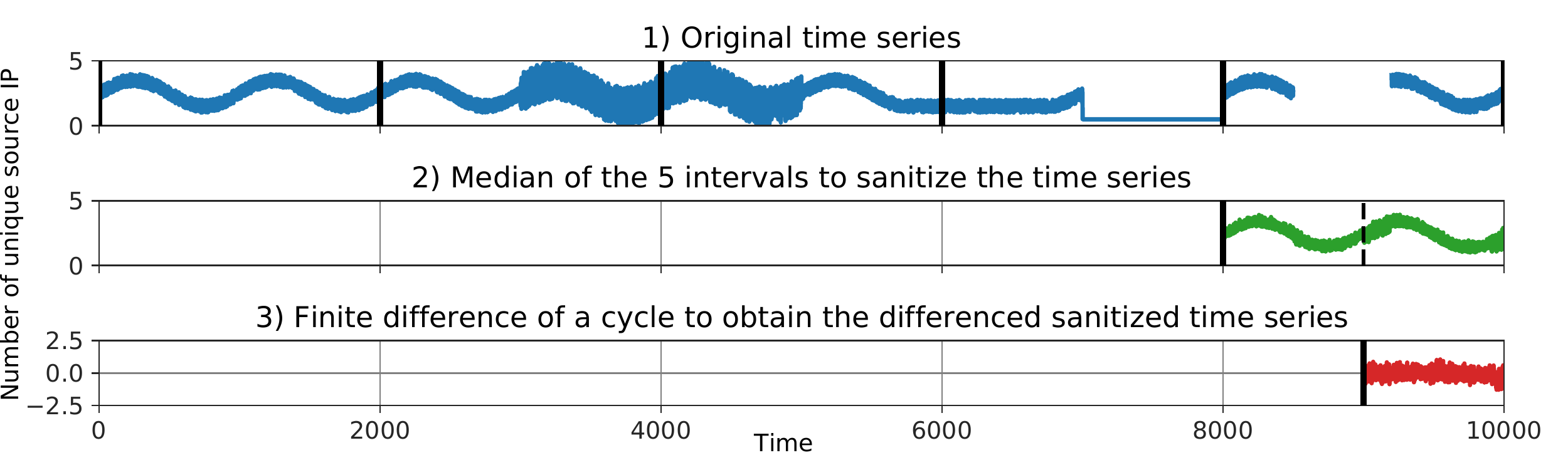}
    \caption{Illustration of preprocessing and seasonal integration of the
    training data
    }\label{fig:4t}
\end{figure*}

Traffic from these streams can be summarized in different
ways, the most common being the \textit{number of bytes}, the \textit{number of packets},
and the \textit{number of unique source IP addresses}.
The number of unique source IP addresses~\cite{dainotti2011analysis} is defined
as the number of IP addresses originating from the same location that contact
the network telescope during a given time interval.
It is an adequate metric to study Internet
outages because it counts the number of devices that send traffic at a geographical
or topological location, while abstracting the need to analyze traffic.
In the event of an outage, some of these devices get disconnected from the Internet, so we expect
to observe drops in the number of unique source IP addresses observed by
the network telescope.

The usage of IBR to detect outages is particularly pertinent
since it is pervasive.
Indeed, the amount of IBR packets that reaches network telescopes is considerable,
incessant, and originates from a variety of
applications~\cite{wustrow2010internet}.
In~\cite{benson2015leveraging}, Benson et al.~performed a spatial
analysis and determined that IBR provided an Internet-wide
view.
All countries, except for 3 with a population of less than
4000 inhabitants, and more than half of all ASs are observed in
their dataset.
Note that half of the ASs that do not show up in the dataset are small,
as they only advertise a /24 prefix, while 86\% of ASs that advertise the
equivalent of a /16
or more are visible. A fifth of the remaining 14\% that do not
generate IBR traffic are blocks that belong to the US government.
The temporal analysis in~\cite{benson2015leveraging} also shows that
most networks frequently generate IBR traffic, in particular when considering
coarse grain aggregations.
Indeed, the median time between observations is shorter than 1 minute for over 90\%
of countries, and is shorter than 10 minutes for about 75\% of the ASs.

To summarize, IBR traffic is ubiquitous, and thus can be used
to detect and analyze large-scale network events. It is continually
sent by a variety of sources all around the world, which makes it a suitable
source to make opportunistic worldwide Internet measurements and specifically
for efficiently detecting outages.

\section{Methodology}\label{sec:method}

In this section, we describe how \name~forecasts the number of unique IP
addresses in IBR traffic and detects outages.
Among the numerous approaches available to forecast time series,
Autoregressive Integrated Moving Average (ARIMA) models are a popular choice
thanks to their simplicity and efficiency~\cite{yu2016improved}.
For this study we select Seasonal-ARIMA (SARIMA)~\cite{gooijer:ijf06} models
in order to deal with weekly patterns observed in IBR time series.
We propose an outage detection method composed of four main steps.
First, we sanitize the training part of the dataset (Section~\ref{sec:pre}) and
we eliminate
non-stationarity in the data by differencing the data with a lag of one week
(Section~\ref{sec:si}).
Second, we compare results with multiple sets of parameters to find the best
parameters for modeling each time series, and compute the corresponding
prediction intervals
(Section~\ref{sec:mod}).
Finally, we detect and report outages based on the differences between the
computed predictions and the actual data (Section~\ref{sec:ana}).

\subsection{Data preparation}\label{sec:pre}
In the following, the IBR time series are split into three sets:
\textit{training}, \textit{calibration}, and \textit{test}. These are used
differently for the modeling (Section~\ref{sec:mod}) and detection phases
(Section~\ref{sec:ana}).
The training and calibration sets are used for the modeling, \ie to learn the
best set of parameters for the ARMA model.
These parameters are then used on the test set to detect potential
outages.

The training data is used as the basis of the predictive model, and we need
to sanitize it.
There are three problems that need to be addressed:

\begin{itemize}
    \item Missing values that we need to fill to have a working model,
    \item Extreme values, which will bias the model by greatly influencing the
        statistical properties of the time series,
    \item The presence of an outage inside the training data, which will lead
        to a model considering outages as the norm.
\end{itemize}

To overcome these problems we assume that the occurrence of missing and extreme
values are uncommon so we can synthesize ten weeks of data into two weeks of
sanitized data.
Our data preparation process is illustrated in Figure~\ref{fig:4t}, with a time
series having the three problems mentioned above.
We consider five intervals of two weeks (top plot in Figure~\ref{fig:4t}),
and compute the median values across all five intervals to obtain two weeks of 
clean synthesized data
(middle plot in Figure~\ref{fig:4t}).
This sanitized time series is then used as the training set for our SARIMA
model.

\subsection{SI\@: Seasonal Integration}\label{sec:si}

ARMA models assume that the data is \emph{stationary},
that is, the  statistical properties of the data (\eg mean and variance) are constant over time.
Because of the strong daily and weekly patterns present in IBR data,
our time series are non-stationary (\eg there
is less traffic at night and during weekends, because more devices are
turned off or disconnected during these periods of time~\cite{internet-sleeps}).
This is the reason why a simple predictive model would not work
with IBR time series.
As a result, we make our time series stationary by filtering these trends
with seasonal differencing.
In practice, our time series contain a weekly and daily trend which we
both remove by applying a seasonal differencing (SI part of SARIMA) of a week
(\eg bottom plot in Figure~\ref{fig:4t}).

The computed training data, which is now sanitized and stationary, can then be
used in the following step to create a predictive model and to make
predictions on the calibration data.

\subsection{ARMA\@: Autoregressive Moving Average}\label{sec:mod}

\begin{figure*}[!ht]
    \centering
    \includegraphics[width=\textwidth]{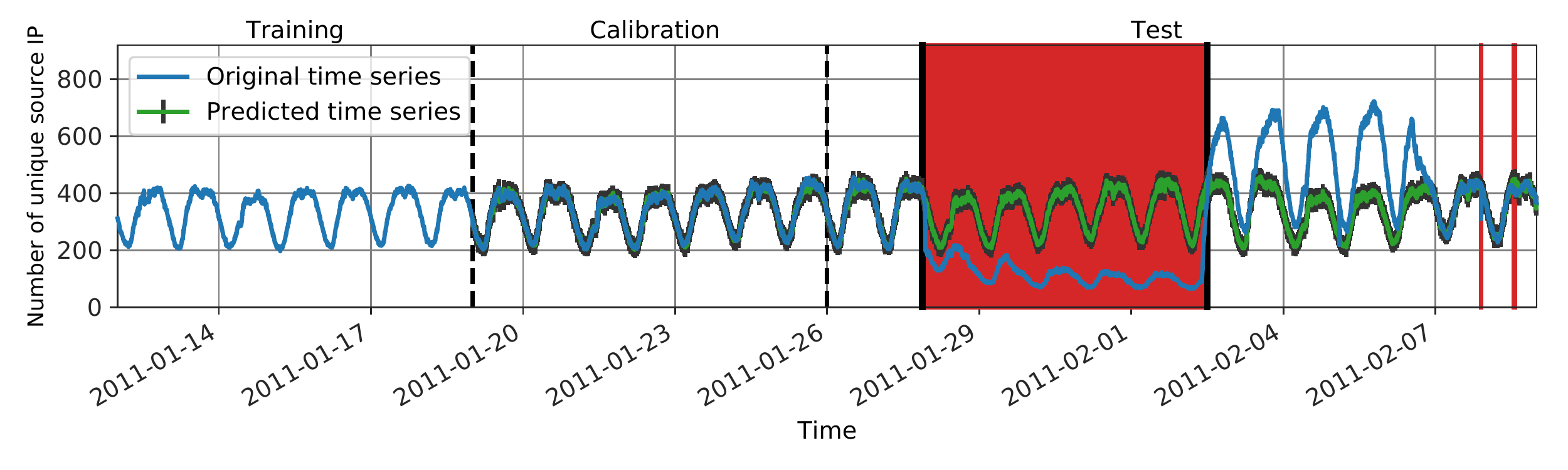}
    \caption{Analyzing the test set with the best model ($AR=4,MA=1$)
    }\label{fig:egypt_outages}
\end{figure*}

In this step, we estimate the best parameters for any given time series.
In pratice, \name~will compute a different set of parameters for each analyzed
time series, which will increase the adaptability of the solution and the
quality of the predictions.
To achieve this goal, we have to precisely estimate the values for the two
key parameters of ARMA, that is the order of the autoregressive model (named
$p$), and the order of the moving-average model (named $q$).
We use the sanitized training data for that purpose, as ARMA models only work
on the condition that the training data is anomaly free and stationary.
In order to find the best combination of parameters for any given time series,
we make predictions on a second set of data that we refer to as the
calibration data.
In practice, we use the period following the training data for defining such a calibration.
We consider several predictive models, each with their own set of $(p,q)$
parameters, to evaluate the performances of various distinct predictions.
We finally compare the accuracy of these predictive models on the data used for this calibration.
We use the \textit{Root Mean Square Error} (RMSE) to compute the error between
the real time series and the one obtained from the predictive model.
We chose the RMSE in order to penalize predictive models that made predictions
that are significantly far from the actual data.
The predictive model (\ie the set of $(p, q)$ parameters) with the lowest
error will thus be used for future predictions.

Now that we have the best parameters to use within the ARMA model, we also 
compute a prediction interval. It defines the boundaries of our predictions and
it is used for the outage detection process. We compute $99.5$\%
prediction intervals using the residual variance. We compute the residual variance
using the \textit{Median Absolute Deviation} (MAD), a robust measure of
data variability used for anomaly detection~\cite{FontugneIMC2018} (the RMSE
being not suitable enough in this case). This is essential, as the
prediction intervals should be both robust to false positives but still
able to capture extreme values introduced by measurement errors and outages.

The model, and its associated prediction interval, are then used to
detect outages, as described in the next section.

\subsection{Detection}\label{sec:ana}

The steps described above provide us with stationary data and an optimized
predictive model for each time series.
The next step is to detect outages with the predictive models.
We define an outage as a point in time where a value of a time series is
smaller than the lower bound of the prediction interval.
The severity of this alarm will be determined by computing the following
distance: \[d = (\hat{X} - X) / (\hat{X} - L),\] where $\hat{X}$ is the predicted
value, $X$ is the actual value from the time series, and $L$ is the
lower bound of the prediction interval.
Distances $d>1$ and $d<-1$ mean that the time series
is outside of the prediction interval,
whereas the time series is within the prediction interval when
$-1 \leq d \geq 1$.
The only cases that are reported as outages are cases
where $d > 1$, that is, when the actual values are outside of the
prediction interval and are smaller than the lower bound
of the prediction interval, which translates in a significant
drop in the number of IPs observed in the time series.
Cases where $d < -1$ (\ie points that are greater than the upper bound of the
prediction interval) are considered as extreme values, but they do not fall
into our definition of an outage, and are thus not reported.

Every hour (\ie 12 data points) we make predictions for the next hour and
compare the actual data to these predictions as explained above.
Each time we move forward in the data, ARMA takes into account the new data points
for the future predictions.
However, we take particular precautions to maintain the quality of the
predictive model.
Data identified as part of an outage should not be used for
future predictions, which brings us back to the problems discussed in Section~\ref{sec:pre}, where
missing values, extreme values, and outages would diminish the quality of the
predictive model.
In this phase, we solve these problems differently, by doing what we refer to as
\textit{inpainting}:
if a new sample of data is considered to be an extreme value (\ie $d < -1$
or $d > 1$), we feed the predictive model with the predicted value instead of
the real one.

\section{Case study}\label{sec:case-studies}

To illustrate the functioning of the proposed method and some of its benefits,
this section provides thorough results from a specific case study.

On January 25th 2011, the Mubarak regime ordered network operators to shut down
Internet connectivity during the
Egyptian revolution, in an attempt to silence the opposition.
The chronology of this event has been described
in~\cite{dainotti2011analysis}.
The authors used BGP routing data, ping, traceroute, and IBR
data. The IBR data was manually analyzed to shed light on the massive 
packet-filtering mechanisms that were put in
place, and to identify denial-of-service attacks related to the political
events happening in Egypt during the same period.
In this section, we present how our solution analyzes the same IBR data but
allows us to systematically detect the beginning and the end of the connectivity
loss, and to estimate the severity of the outage.

Figure~\ref{fig:egypt_outages} shows 
the time series of unique source IP addresses from Egypt reaching the UCSD
Network Telescope (plotted in blue).
The disconnections occurred between the 28th of January and the 3rd of February,
2011,
as it can be seen by the loss of intensity of the signal depicted in the figure.
Here, we chose to include in our analysis also the values of the time series
after the outages,
because of an interesting phenomenon that was occurring:
the values of the time series are higher than usual during the days
that follow the Egyptian revolution and
go back to normal around the 7th of February.
In~\cite{sipscan}, the authors revealed that a botnet covertly (and massively) scanned
the Internet during those days.

This time series is analyzed as follows.
The training set, to the left, is sanitized following the methods discussed
in~\ref{sec:pre}.
Multiple sets of ARMA parameters are then going to be used to predict the
calibration set.
The predictions are plotted with a green line.
The set of parameters that resulted in the lowest error ($p=4, q=1$ in this
case) will be used for the rest of the analysis.
The difference between the predicted time series and the original time series
allowed us to
compute prediction intervals using the MAD\@. These intervals are plotted with
gray bars that surround the predictions.

Then the test set is compared to the ARMA model and the prediction intervals
computed in the previous step.  The sudden drop that occurs when the outage
starts, puts the time series below the prediction intervals, which means that an
outage is reported.  Visually, this is shown with a red vertical line.
Additionally, it also means that the \textit{inpainting} process described in
Section~\ref{sec:ana} will take place, which is clear here, since the trend of
the predicted time series stays similar to that of the original time series,
even if an outage is occurring at the same time.  No alarm is reported during
the botnet activity (\cite{sipscan}) that follows the outage, because the
original time series values are higher than our prediction intervals, which
means that the data is again inpainted and it will not count as an anomaly.

\section{Validation, Calibration and Comparison}\label{sec:validation}

We evaluate the limits, and performance of \name~through a validation and a comparison.
We start by considering a set of verified outages from our ground-truth dataset,
which we use to assess the accuracy of our
outage detector, and look for the best threshold, \eg the one determining the minimal number of IPs required to make accurate predictions.
We then use a different set of outages in order to compare
\name~against CAIDA's outage detection techniques (using BGP dumps, active probing and the network telescope
data).

\subsection{Validation}\label{sec:val}

In this section, we evaluate the reliability of our technique using a
reference dataset and gathering 130 time series containing
outages. These time series contain three different types of spatial
aggregates---ASs, countries, and regions within countries---from various years
(2009 to
2018). The duration of these outages spans from an hour to a week.
The comprehensive list of time series that compose this dataset is given
in Table~\ref{tab:ground_truth}.
As an example, the RIPE NCC and Duke University BGP experiment~\cite{ripeduke}
caused several outages in different ASs worldwide by triggering a bug in some
Cisco routers.

\begin{table}[b]
    \caption{Number of time series per IP threshold and per spatial scale}
    \centering\label{fig:ts_number}
\resizebox{\columnwidth}{!}{%
\begin{tabular}{@{}lccccc@{}}
\toprule
 & $>10$ & $>15$ & $>20$ & $>25$ & Total \\ \midrule
Countries & \begin{tabular}[c]{@{}c@{}}144\\ (56.9\%)\end{tabular} & \begin{tabular}[c]{@{}c@{}}135\\ (53.3\%)\end{tabular} & \begin{tabular}[c]{@{}c@{}}128\\ (50.5\%)\end{tabular} & \begin{tabular}[c]{@{}c@{}}120\\ (47.3\%)\end{tabular} & 253 \\
Regions & \begin{tabular}[c]{@{}c@{}}1,038\\ (21.4\%)\end{tabular} & \begin{tabular}[c]{@{}c@{}}879\\ (18.1\%)\end{tabular} & \begin{tabular}[c]{@{}c@{}}778\\ (16.0\%)\end{tabular} & \begin{tabular}[c]{@{}c@{}}704\\ (14.5\%)\end{tabular} & 4,846 \\
ASs & \begin{tabular}[c]{@{}c@{}}1,157\\ (1.8\%)\end{tabular} & \begin{tabular}[c]{@{}c@{}}867\\ (1.4\%)\end{tabular} & \begin{tabular}[c]{@{}c@{}}719\\ (1.1\%)\end{tabular} & \begin{tabular}[c]{@{}c@{}}621\\ (1.0\%)\end{tabular} & 61,639 \\ \bottomrule
\end{tabular}%
}
\end{table}

\begin{figure}[htb]
    \includegraphics[width=\columnwidth]{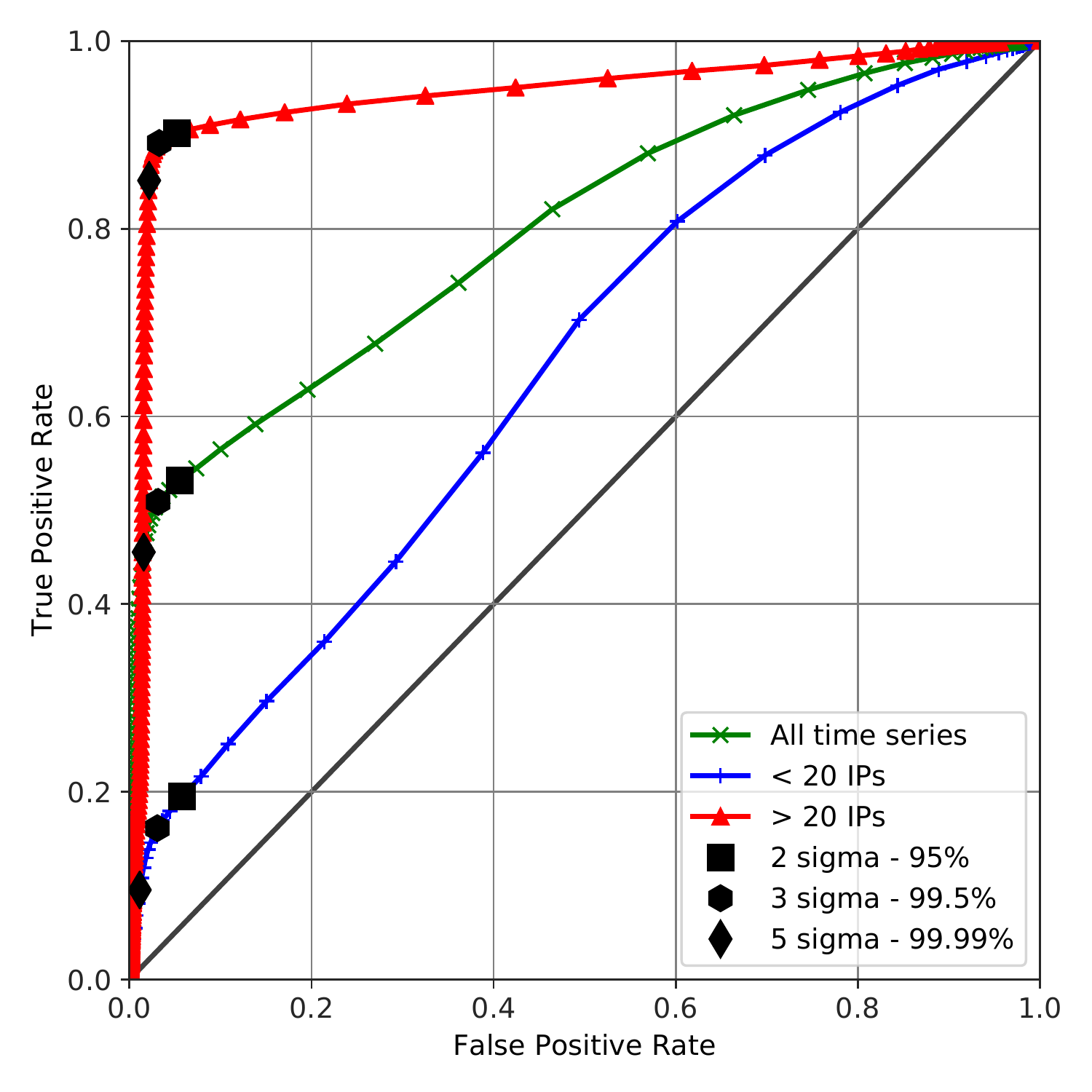}
    \caption{ROC curve with a 5 minutes time grain and a
    threshold of 20}\label{fig:ip20}
\end{figure}

We evaluate \name~by computing the True Positive Rate (TPR) and the
False Positive Rate (FPR), and show our calibration results with a ROC curve.
Our purpose is twofold: we look into the accuracy of our approach, and we
search for its best parameters by exploring its calibration spectrum.
In particular, we determine which confidence level should be used to
assess whether an outage is occurring or not.
Our aim is to find the best trade-off between the TPR and the FPR by
considering our collection of documented outages as the ground truth.

Moreover, to quantify the ability of our method to maximize the TPR while
keeping the FPR low, we need to set two evaluation parameters used in our ROC
analysis. On the one hand, we need to find out the minimal intensity
required in the time series for our method to finely operate, and on the other
hand, the smallest time granularity at which we can accurately detect outages.
The intensity of time series is measured as the median number of observed IPs
in a week.
Trying multiple thresholds showed us that \name~yielded better results with a
threshold of 20 IPs, and that increasing this number had little effect on the
accuracy. The results are presented in Figure~\ref{fig:ip20}, where three
different ROC curves are plotted:

\begin{itemize}
    \item The green curve plots the accuracy for all time
        series;
    \item The red curve plots the accuracy for time series with a
        median of IP addresses in a week that is greater than 20;
    \item The blue curve plots the accuracy for time series
        with a median of IP addresses in a week that is smaller than 20.
\end{itemize}

On the one hand, using a small number of IP addresses provides performance only
slightly better than
using a random model, which is expected, as the central limit theorem does not
hold for samples that are too small.
On the other hand, the higher the number of IPs is, the better the performance.
(the red curve yields much better results than the blue one).
The accuracy of our method for all time series (the green curve), is not
satisfactory because of the influence of the time series contained in the blue
curve.
As a result, we have chosen to limit our analysis to the time series that have a
median of more than 20 IPs per week. 

Table~\ref{fig:ts_number} summarizes the impact of this threshold on the
number of remaining time series.
Setting this threshold to 20 limits the number of time series that we can analyze
to 1625, but it significantly increases the accuracy of our detector.
Here, we make the assumption that network operators will want to have a low
FPR, even if it means missing smaller outages.
We also found that the size of the time bins we use can be relatively small (around 5 minutes)
without impacting the performance much.
This analysis is not included due to space constraints.

To conclude this section, we recommend to use a threshold of 20 IPs for the time
series and 5 minutes long time-windows as in Figure~\ref{fig:ip20}. These two
parameters can of course be tuned according to the data collection's
specificity. Using such a threshold and time granularity (we can estimate outage
durations at a 5 min granularity), the best confidence level for the prediction
intervals is $99,5\%$ ($3\sigma$).
With these settings we obtain an acceptable true positive rate of $90\%$ while
keeping the false positive rate under $2\%$.

\subsection{Comparison}\label{sec:comp}

\begin{figure*}[ht]
    \centering%
    \begin{subfigure}{.49\textwidth}
        \centering%
        \includegraphics[width=\textwidth]{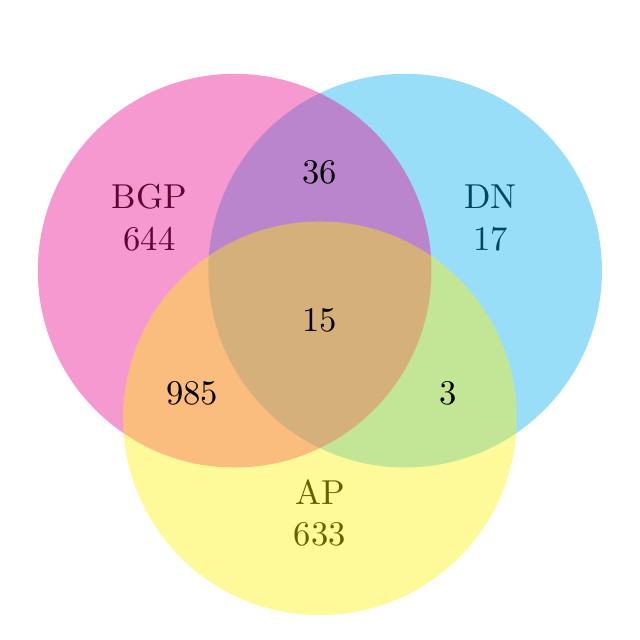}
        \caption{BGP IODA versus AP IODA versus DN IODA}\label{fig:c1}
    \end{subfigure}
    \begin{subfigure}{.49\textwidth}
        \centering%
        \includegraphics[width=\textwidth]{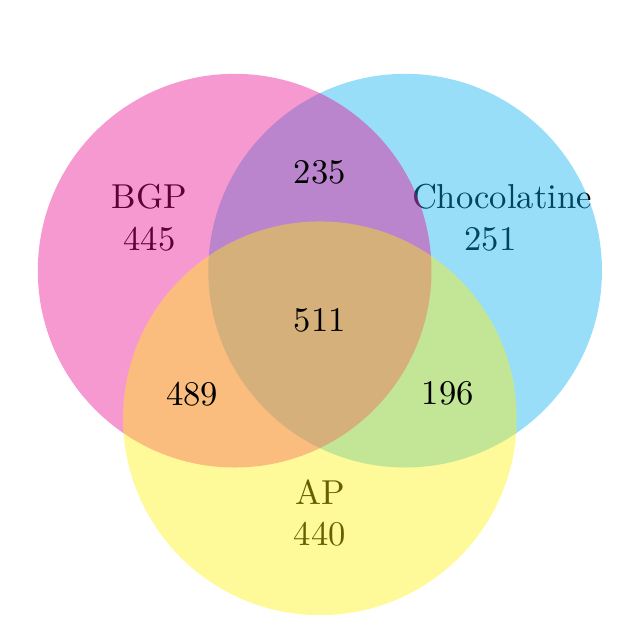}
        \caption{\name~versus BGP IODA versus AP IODA}\label{fig:c2}
    \end{subfigure}
    \caption{Comparison of the number of 5 minutes time bins identified as
    outage per detector}\label{fig:comp}
\end{figure*}

In this section, we compare the performance of our detector to three other
techniques hosted in IODA: CAIDA's darknet detector (DN), CAIDA's BGP detector
(BGP), and a technique based on active probing (AP), Trinocular~\cite{quan2013trinocular}. A description of
the integration of these 3 detectors in IODA can be found in~\cite{ioda-help}.
In order to compare the detectors, we use a second ground-truth
sample to emphasize the versatility of \name~on different time series.
Its set of outages is distinct from the previous one, but still
decomposed in 5 minutes bins (see Table~\ref{tab:ground_truth2}). We ran the
4 detectors and enumerated the number of 5
minutes time bins where an outage is detected, for each detector.
Fig.~\ref{fig:comp}~\subref{fig:c1} plots the number of outages detected by
IODA's components, and Fig.~\ref{fig:comp}~\subref{fig:c2} plots how
\name~compares against BGP and Active Probing (AP) detectors.
Note that the number of events given below the name of each detector are events
detected only with that technique.
The intersections depict the number of events detected by multiple detectors.
For example, there are 1680 BGP events (the sum of each intersection
combination in the magenta based
set), 985 of which are also detected by the active probing technique.

Comparing \name~with the IODA's darknet detector, one can observe that
\name~detects two order of magnitude more outages, 1193 compared to 71.
This result highlights the much higher sensitivity of our approach, while
CAIDA's darknet detector is extremely conservative by nature.
By modeling weekly, and \textit{a fortiori} daily, patterns our predictions are
adaptively following the time series oscillations, while this is not
the case in CAIDA's detector, which uses a global threshold approach.


Another way to evaluate \name~is to cross-reference the set of alarms it is
able to detect compared to the other detectors (and look at all the
intersections). When there are intersections, the corresponding events
are very likely to be actual outages, \ie they are true
positives. Fig.~\ref{fig:comp}~\subref{fig:c2} shows that the outages detected
by \name~are likely to intersect the outages of the other sources. Indeed, there
are only 251 alarms that are specific to \name. The analysis of these alarms
shows us that 59\% of them occur in a range of 1 hour around alarms detected by
other data sources.
Generally speaking, these results suggest that our tool is complementary to
the two others (BGP and AP) and clearly outperforms IODA's current darknet
detector.

\section{Reproducibility}\label{sec:repro}

All results presented in this paper are easily reproducible by the research
community.
Our source code is made publicly available~\cite{guillot} and it
automatically fetches and processes IBR data, which
means that the dataset is also available.
The code is structured in such a way that one simply needs to format its
data according to our data format to be able to launch \name~on
different data sources.

\section{Conclusion}\label{sec:conclusion}

In this paper we proposed \name, which detects remote outages using Internet
Background Radiation traffic. The underlying predictive methodology is based on SARIMA models.
Both the method and the data are easy to respectively deploy and collect in most ISP.
We show that our method detects outages as quickly as 5 minutes after their
occurrence, with a $90\%$ true positive rate and a small percentage of false
alarms ($2\%$).
\name~is able to detect outages in time series with as little as 20 IP addresses.
Moreover, we compare its performance against other passive and active detectors.
Its share of common events with other sources is high, while each technique seems able to reveal specific events.
We observe that the shares of common events, the overall and two-by-two intersections, are the most significant, while each technique seems able to reveal specific events too.

Our method is tailored to seasonal data and is robust to noise. It is
therefore applicable to many other data sources reflecting Internet activity. For
example, we plan to experiment its deployment on access logs of widely popular
content, while its operational integration into the CAIDA's IODA outage detection
system~\cite{ioda} is already in progress.


\begin{table*}[t]
    \caption{Ground truth --- validation (Section~\ref{sec:val})}
    \centering\label{tab:ground_truth}
    \begin{tabularx}{\textwidth}{@{}llX@{}}
\toprule
Event & Detection Time Frame & Time series \\ \midrule
        Czech ISP & 16--02--2009 16:20--17:20 & \scriptsize{AS=\{62, 135, 158, 166,
        223, 291, 348\}} \\
        AfNOG & 03--05--2009 12:00--13:00 & \scriptsize{AS=\{3, 242, 467\}} \\
CNCI & 17--08--2009 18:00--18:40 & \scriptsize{AS=\{80, 149, 333, 360, 524,
        580, 585\}} \\
RIPE-Duke & 27--08--2010 08:30--09:30 & \scriptsize{AS=\{48, 54, 56, 63, 95,
        143, 153, 204, 209, 210, 283, 310, 374, 377, 384, 385, 397, 398, 443,
        474, 475, 483, 488, 497, 509, 564, 575, 595, 674, 676, 694, 714, 788,
        791\}} \\
JunOS bug & 07--11--2011 14:00--15:00 & \scriptsize{AS=\{7, 45, 68, 71, 73, 93,
        119, 160, 177, 181, 187, 209, 215, 229, 257, 260, 273, 278, 297, 314,
        316, 317, 320, 322, 324, 325, 328, 332, 335, 336, 337, 347, 392, 414,
        415, 425, 429, 431, 479, 485, 490, 493, 501, 504, 529, 535, 569, 597,
        624, 628, 636, 647, 650, 654, 655, 697\}} \\
Egypt & \begin{tabular}[c]{@{}l@{}}01--27--2011 21:00 --
02--02--2011 12:00\end{tabular} & \scriptsize{Countries=\{Egypt\}, Regions=\{
        978, 971, 984, 985, 974, 979, 980, 993\}, AS=\{8452, 36992, 24863, 24835\}} \\
Brazil & \begin{tabular}[c]{@{}l@{}}21--03--2018 18:45 --
22--03--2018 10:00\end{tabular} & \scriptsize{Countries=\{Brazil\},
        Regions=\{Amazonas, Bahia, Caera, Distrito Federal\}} \\
Syria & \begin{tabular}[c]{@{}l@{}}27--05--2018 22:00 --
28--05--2018 06:00\end{tabular} & \scriptsize{Countries=\{Syria\}} \\
Syria & 30--05--2017 00:00--06:00 & \scriptsize{Countries=\{Syria\}} \\
Azerbaijan & \begin{tabular}[c]{@{}l@{}}02--07--2018 12:00--03--07--2018
18:00\end{tabular} & \scriptsize{Countries=\{Azerbaijan\}} \\
DRC & \begin{tabular}[c]{@{}l@{}}23--12--2017 15:00--26--12--2017
09:00\end{tabular} & \scriptsize{Countries=\{Democratic Republic of the Congo\}} \\
Gambia & \begin{tabular}[c]{@{}l@{}}30--11--2016 17:00--04--12--2016
22:00\end{tabular} & \scriptsize{Countries=\{Gambia\}} \\ \bottomrule
\end{tabularx}
\end{table*}

\begin{table}[t]
    \caption{Ground truth --- comparison (Section~\ref{sec:comp})}
    \centering\label{tab:ground_truth2}
    \begin{tabularx}{.49\textwidth}{@{}llX@{}}
\toprule
Event & Detection Time Frame & Time series \\ \midrule
Angola & \begin{tabular}[c]{@{}l@{}}07--09--2018 16:57 -- 08--09--2018
06:20\end{tabular} & \scriptsize{Countries=\{Angola\}} \\
Iraq & \begin{tabular}[c]{@{}l@{}}13--10--2018 15:10 -- 18--10--2018
19:32\end{tabular} & \scriptsize{Countries=\{Iraq\}} \\
Venezuela & \begin{tabular}[c]{@{}l@{}}15--10--2018 18:00 -- 19--10--2018
05:00\end{tabular} & \scriptsize{Countries=\{Venezuela\}} \\
Tajikistan & 26--10--2018 10:00--14:20 & \scriptsize{Countries=\{Tajikistan\}} \\
Ivory Coast & \begin{tabular}[c]{@{}l@{}}28--10--2018 23:00 -- 29--10--2018
08:00\end{tabular} & \scriptsize{Countries=\{CI\}} \\
Argentina & \begin{tabular}[c]{@{}l@{}}17--11--2018 11:00 -- 18-11-2018
01:00\end{tabular} & \scriptsize{Countries=\{Argentina\}} \\
Syria & \begin{tabular}[c]{@{}l@{}}18--11--2018 22:00 -- 19--11--2018
03:00\end{tabular} & \scriptsize{Countries=\{Syria\}} \\
Taiwan & 19--11--2018 00:00--06:00 & \scriptsize{Countries=\{Taiwan\}} \\
Armenia & 20--11--2018 11:00--15:00 & \scriptsize{Countries=\{Armenia\}} \\
Algeria & 30--11--2018 03:00--17:00 & \scriptsize{Countries=\{Algeria\}} \\
Gabon & 11--12--2018 17:00--23:55 & \scriptsize{Countries=\{Gabon\}} \\
Kyrgyzstan & \begin{tabular}[c]{@{}l@{}}11--12--2018 22:00 -- 12--12--2018
02:00\end{tabular} & \scriptsize{Countries=\{Kyrgyzstan\}} \\
AS 209 & \begin{tabular}[c]{@{}l@{}}27--12--2019 15:00 -- 28--12--2018 01:00\end{tabular} & AS=\{209\} \\
Ethiopia & 03--01--2019 11:00--15:00 & \scriptsize{Countries=\{Ethiopia\}} \\
Cameroon & \begin{tabular}[c]{@{}l@{}}14--01--2019 11:00 -- 15--01--2019
10:00\end{tabular} & \scriptsize{Countries=\{Cameroon\}} \\
Indonesia & \begin{tabular}[c]{@{}l@{}}14--01--2019 05:00 -- 15--01--2019
08:00\end{tabular} & \scriptsize{Countries=\{Indonesia\}} \\
Zimbabwe & \begin{tabular}[c]{@{}l@{}}15--01--2019 04:00 -- 17--01--2019
12:00\end{tabular} & \scriptsize{Countries=\{Zimbabwe\}} \\
Zimbabwe & \begin{tabular}[c]{@{}l@{}}17--01--2019 20:00 -- 18--01--2019
16:00\end{tabular} & \scriptsize{Countries=\{Zimbabwe\}} \\
Panama & \begin{tabular}[c]{@{}l@{}}20--01--2019 15:00 -- 21--01--2019
01:00\end{tabular} & \scriptsize{Countries=\{Panama\}} \\
Laos & 24--01--2019 17:00--21:00 & \scriptsize{Countries=\{Laos\}} \\
Panama & 29--01--2019 13:00--23:55 & \scriptsize{Countries=\{Panama\}} \\
Morocco & 11--02--2019 06:00--16:00 & \scriptsize{Countries=\{Morocco\}} \\\bottomrule
\end{tabularx}
\end{table}

\section*{Acknowledgments}

The authors thank Brandon Foubert, Julian Del fiore, and Kenjiro Cho for
their valuable comments.
This work has been partially funded by the IIJ-II summer internship program,
and has been made possible in part by a grant from the Cisco
University Research Program Fund, an advised fund of Silicon Valley Foundation.
This research is supported by the National Science Foundation grant
CNS-1730661, by the U.S. Department of Homeland Security S\&T
Directorate via contract number 70RSAT18CB0000015, by the Air Force Research
Laboratory under agreement number FA8750-18-2-0049, and by the Open Technology Fund.

\bibliographystyle{splncs04}
\bibliography{\jobname}

\end{document}